\documentclass[]{article}
\usepackage{amsmath}
\usepackage{amssymb}
\usepackage{amsfonts}
\usepackage{graphicx}
\usepackage[margin=1.0in]{geometry}
\usepackage{amsthm}
\theoremstyle{definition}

\newcommand{\captionfonts}{\small}

\makeatletter  
\long\def\@makecaption#1#2{%
  \vskip\abovecaptionskip
  \sbox\@tempboxa{{\captionfonts #1: #2}}%
  \ifdim \wd\@tempboxa >\hsize
    {\captionfonts #1: #2\par}
  \else
    \hbox to\hsize{\hfil\box\@tempboxa\hfil}%
  \fi
  \vskip\belowcaptionskip}
\makeatother   

\title{Price and Quantity Trajectories: Second-order Dynamics}
\author{Eric Kemp-Benedict\\
Stockholm Environment Institute\\
eric.kemp-benedict@sei-international.org}

\begin{document}

\maketitle
\bibliographystyle{plain}

\begin{abstract}
In two previous papers the author developed a second-order price adjustment (t\^atonnement) process. This paper extends the approach to include both quantity and price adjustments. We demonstrate three results: a analogue to physical energy, called ``activity'' arises naturally in the model, and is not conserved in general; price and quantity trajectories must either end at a local minimum of a scalar potential or circulate endlessly; and disturbances into a subspace of substitutable commodities decay over time. From this we argue, although we do not prove, that the model features global stability, combined with local instability, a characteristic of many real markets. Following these observations and a brief survey of empirical results for price-setting and consumption behavior in markets for ``real'' goods (as opposed to financial markets), we conjecture that Stigler and Becker's well-known theory of consumer preference opens the possibility of substantial degeneracy in commodity space, and therefore that price and quantity trajectories could lie on a relatively low-dimensional subspace within the full commodity space.

\vspace{2em}

\noindent\textit{Keywords: disequilibrium economics; Walrasian adjustment; Marshallian adjustment; SMD theorem; Helmholtz-Hodge decomposition theorem}
\end{abstract}

\section{Introduction}
Two extreme views dominate the literature on market adjustment mechanisms, one associated with new Keynesian economics and the other with the rational expectations hypothesis. In the first, prices are considered ``sticky'', while quantities respond comparatively rapidly to restore equilibrium; in the second, prices adjust quickly enough that the adjustment process itself can effectively be ignored \cite{flaschel_dynamic_1997}. Of these two views, the second is more generally accepted. A belief that markets adjust rapidly to clear prices justifies both the General Equilibrium models that inform much practical economic policy (e.g., \cite{ginsburgh_structure_2002}) and the efficient markets hypothesis \cite{fama_efficient_1970,malkiel_efficient_2003} that underlies such practical financial tools as the capital asset pricing model \cite{sharpe_capital_1964} and the Black-Scholes equation \cite{black_pricing_1973}.

Indeed, some presentations of General Equilibrium theory go farther than claiming merely that prices clear extremely rapidly. They argue instead that equilibrium conditions cannot result from a dynamic process \cite{ginsburgh_structure_2002}, and any dynamic behavior---key economic quantities changing over time---arises from sequential equilibria determined by external changes \cite{heer_dynamic_2009}. This startling claim arises from two equally startling results about price adjustment mechanisms: the SMD theorem and the ubiquity of instability. The SMD theorem, elaborated by Sonnenschein \cite{sonnenschein_market_1972}, Mantel \cite{mantel_characterization_1974}, and Debreu \cite{debreu_excess_1974} shows that almost any continuous function of prices can be expressed as the sum of well-behaved individual demand and supply functions, so that an aggregate excess demand function can take almost any form. Even more troubling, the individual demand and supply functions can be arbitrarily close to each other and still produce this result, so that small changes in individual demand and supply behavior can result in large  changes in aggregate excess demand. Moreover, Saari \cite{saari_iterative_1985} showed that even with a fixed aggregate excess demand function, any (first-order) iterative price mechanism that uses only local information cannot be guaranteed to reach a price equilibrium in a large set of well-behaved economies; thus, instabilities are ubiquitous.

These theorems are widely accepted by both supporters \cite{ginsburgh_structure_2002} and critics \cite{ackerman_still_2001} of General Equilibrium, but not uniformly. Balasko \cite{balasko_equilibrium_2009} argues that the SMD result is more narrow than generally believed, and Saari's instability result has not stopped continued research into disequilibrium dynamic models \cite{flaschel_dynamic_1997,herings_equilibrium_1997,keisler_getting_1996}. Because setting prices, reacting to prices, making and purchasing goods, and deciding to hold goods as stocks or to release them are the visible manifestations of day-to-day economic life, the current situation is unsatisfactory. In this paper we take the position that the pervasive instabilities of existing theories of economic dynamics is a problem with the theory that should not lead us to conclude that real economies have no dynamics beyond a changing sequence of equilibria responding to external shocks. Indeed, a theory of dynamics is essential to equilibrium economics, because the equilibria are not meaningful if they are unstable, and stability cannot be proven without dynamics \cite{fisher_disequilibrium_1983}.

In this paper we propose a mechanism for quantity and price adjustment that is second-order in time, rather than the conventional first-order ``proportional-control'' adjustment mechanism.\footnote{This paper is self-contained, but builds on two previous papers \cite{kemp-benedict_second-order_2011,kemp-benedict_second-order_2012}. It partly repeats, and partly extends, material from those papers.} It exhibits what Fisher \cite{fisher_disequilibrium_1983} called ``favorable suprises,'' in which the dynamics produce recurring opportunities for arbitrage. The model we propose is similar to the derivative-control adjustment mechanism of Flaschel, Franke, and Semmler \cite{flaschel_dynamic_1997}, but unlike their mechanism, which was introduced specifically to improve convergence, the mechanism proposed in this paper arises naturally in economies with heterogeneous agents, some following first-order price dynamics and the rest waiting for others to act first. We argue (but do not prove) that the continuous-time version of the theory can lead to global stability combined with local instability. We also show that an analogue to physical energy arises naturally in the theory. This quantity is not conserved, and the non-conservative terms have compelling interpretations.

\section{Empirical Findings on Price Setting and Consumer Responses}
Before descending into the very formal model presented in this paper we first step back and survey the empirical literature. Our interest is in markets for real goods, in contrast to financial markets. The actors in the markets of interest populate the entire production chain, from the production of raw commodities (crude oil, crops, metal ore) through processing and manufacturing industries, to wholesalers, retailers, and the final consumer. In this section we review empirical findings on price-setting behavior by firms and retailers, and how consumers react to retailer's pricing strategies. All of the studies use data from North America, so they may reflect cultural or behavioral patterns specific to the region.

\subsection{Price setting by firms}
Over the relatively long time scale of the lifetime of a product, prices gradually decline and converge. In a study of consumer durables, Curry and Riesz \cite{curry_prices_1988} examined Consumer Reports data over a 10-year period, and observed both declines and convergence as products matured. The convergence did not happen rapidly and was more consistent with changes in strategy, quality, and price across a product's life cycle than with active price setting in a dynamic market.

Also over a relatively long time scale, but in this case over the lifetime of a brand, firms can influence relative prices even for very similar products through advertising. Willis and Mueller \cite{wills_brand_1989} used national wholesaler price data for relatively homogeneous food products for the US. They argued that since most wholesalers apply a uniform markup to firm prices, differences in wholesale prices match those of firm prices. They found that price was positively associated with advertising expenditure, and concluded that for homogeneous goods, market power is the most important determinant of price premiums and higher profits.

Blinder \cite{blinder_why_1991,blinder_sticky_1994,blinder_asking_1998} focused on the shorter-term price-setting behavior of firms. He and his team carried out a survey of randomly-selected for-profit unregulated firms in the private non-farm sector that asked about price-setting behavior. Less than two per cent of the sample set prices more than once a day; the modal value was once a year; and almost 80 per cent set prices four times or fewer each year. The results showed that firms' prices are indeed ``sticky'' in that they rarely change. When managers were asked why they did not change prices more frequently, the most common responses were that it would be problematic for their customers; they faced competition; the costs of changing prices were prohibitive; and their costs did not change more frequently than they changed prices.

The survey also asked managers what led them to change prices and to set a particular price level. Of the twelve strategies that the survey explored, four were identified by at least one-half of the surveyed firms as ``moderately important'' or higher. In decreasing rank order, they were:
\begin{enumerate}
\item Coordination failure: Firms hold back on price changes, waiting for other firms to go first (mainly because of a fear of losing sales if they raised prices first).
\item Cost-based pricing with lags: Firms delay a rise in price until costs rise.
\item Delivery lags, service, etc.: Firms prefer to vary other factors, such as delivery lags, service, or product quality.
\item Implicit contracts: Firms tacitly agree to stabilize prices, perhaps out of ``fairness'' to customers.
\end{enumerate}
These strategies are clearly far from the behavior that the rational expectations hypothesis envisions, and is more consistent with behavioral approaches to economic activity \cite{simon_behavioral_1984,simon_models_1997}.

\subsection{Price setting by retailers}
Retailers change prices much more frequently than do manufacturing firms or wholesalers, and different sectors develop their own patterns of price-setting behavior. For example, a study of gasoline prices in Vancouver showed that prices went through a roughly weekly cycle, with price rises occurring on Tuesdays or Wednesdays. Price rises happened rapidly and in a spatially homogeneous manner and then decayed slowly at different rates in different parts of the city \cite{eckert_retail_2004}.

Store managers adopt pricing strategies that provide them with heuristic rules for setting prices. Bolton and Shankar identified five such strategies, including the frequently studied everyday low pricing (EDLP) and high-low (Hi-Lo) strategies \cite{bolton_empirically_2003}. Also, specific types of products, such as store brands and seasonal goods, are associated with specific pricing behavior. Examining store brands in US cities, Bonfrer and Chintagunta \cite{bonfrer_store_2004} found that, when introducing a store brand, retailers tend to mark up incumbent brands. In contrast, when introducing a new national brand, retailers mark down incumbent brands. Considering seasonal goods, MacDonald \cite{macdonald_demand_2000} and Chevalier et al. \cite{chevalier_why_2000} examined the surprising phenomenon that prices for seasonal goods decline during periods of peak demand.  Using data from the US, MacDonald found that the price declines were more pronounced where there was significant competition, and that price declines coincided with increases in promotional activity. Chevalier et al.'s results are consistent with MacDonald's but they also criticize MacDonald for possibly confounding a surge in demand due to lower prices with lower prices being driven by an anticipated seasonal increase in demand. Using a unique data set for a chain of stores in the Chicago area that included both wholesale and retail prices, Chevalier et al. found that price changes were not related to changes in costs and were most consistent with a loss-leader model, in which stores advertise sales in order to draw customers.

\subsection{Consumer response to prices}
There is considerable evidence that people have a mental ``reference price'' against which they judge prices of goods, and they form references prices based on their previous observations of prices. They are also more sensitive to ``losses'' (higher than expected prices) than they are to ``gains'' (lower than expected prices) \cite{kalyanaram_empirical_1995}. Briesch et al. \cite{briesch_comparative_1997} used scanner data for locations in the US to test competing theories of reference prices. They found that for peanut butter, ground coffee, and tissue, the most successful model proposed that consumers keep a memory of price changes for that brand as their reference price, independent of the price trajectories of other brands. For liquid detergent consumers were more likely to compare prices when they visited the store.

Diverse factors influence consumers' decisions. Kaul and Wittink \cite{kaul_empirical_1995}, in a review, found that price advertising contributed to greater price sensitivity on the part of consumers, while non-price advertising reduced price sensitivity. Also, consistent with the results on pricing of seasonal goods, they found that price advertising tends to be associated with lowered prices. Guadagni and Little \cite{guadagni_logit_1983} examined scanner data on coffee purchases in Kansas City, in the US. They fit a multinomial logit model to the data and showed that brand loyalty and package size loyalty were both important factors in purchasing decisions. They also found that price and promotions (either with or without a price reduction) affected purchasing decisions. Han et al. \cite{han_consumer_2001} successfully tested a model of thresholds in price mark-downs, suggesting that consumers' response to price changes is not continuous. Finally, in the study of store brands cited above, Bonfrer and Chintagunta \cite{bonfrer_store_2004} found that consumers who had strong brand loyalty tended to have less store loyalty, and vice-versa, and consumers who had strong store loyalty were more likely to buy the store brand.

\subsection{Discussion}
These results show that price changes occur on different time scales in response to a variety of influences. Some of the patterns can reasonably be treated as perturbations: gasoline prices in Vancouver undergo a rapid but relatively small-amplitude cycle of price setting as retailers repeat a pattern of tacit collusion to raise prices above the equilibrium price, followed by a gradual return to equilibrium. Also, the observed behavior is not inconsistent with price-setting due to excess demand. For example, retailers periodically reduce prices on some products, and except for seasonal price reductions, they could presumably decide to do this for products that are moving slowly. However, many factors enter into both pricing and purchasing decisions, and those decisions rely heavily on heuristics.

These complications should be kept in mind as we now turn to a rather formal model. We introduce heuristic behavior into a conventional t\^atonnement formula, but only one kind of heuristic. We shall show that even this modest addition leads to substantive changes in price and quantity dynamics, raising the possibility that a full model of quantity and price adjustment could look more different still.

\section{First and Second-order Quantity and Price Dynamics}
We start with what Flaschel, Franke, and Semmler \cite{flaschel_dynamic_1997} call a proportional control mechanism with cross-effects. That is, we assume that sellers may adjust either prices, quantities, or both, if they perceive the economy---or at least their part of it---to be out of equilibrium.\footnote{Mechanisms in which only prices changes are sometimes called ``Walrasian'', while mechanisms in which only quantities change are called ``Marshallian''. However, these designations are not entirely accurate \cite{hamouda_expectations_1988}.} For example, a retailer might move a slowly-selling product to a ``bargain bin'' and replace it on the shelf with a more attractive substitute; a worker may demand a raise; a firm may make a quarterly adjustment to the wholesale prices of its major product lines; or a wholesaler may engage in inventory speculation, holding a product in the warehouse if demand is currently low in the expectation that it will later rise. Generally, prices are assumed to react to excess demand, while quantities are assumed to react to prices out of equilibrium. That is,
\begin{subequations}\label{eqn_pdot_qdot}
\begin{align}
\dot p &= \kappa\eta(p,q),\\
\dot q &= \lambda\pi(p,q).
\end{align}
\end{subequations}
In these equations, $p$ and $q$ are vectors of prices and quantities, with one value for each commodity. The functions $\eta(p,q)$ and $\pi(p,q)$ are the excess demand and the gap between current prices and their equilibrium values, which we shall call the price deviation. The coefficients $\kappa$ and $\lambda$, which are assumed to be diagonal matrices, convert gaps in demand or price into rates of change of price or quantity for each commodity. In Equations \ref{eqn_pdot_qdot}, prices and quantities of different commodities in different markets may react quickly or slowly to changes in excess demand and price fluctuations. Also, we do not place any constraints on the equations other than to assume they are continuous and defined over a compact space.\footnote{Although these assumptions are conventional, they are also problematic. Goods are granular, and prices cannot take on any real value \cite{nadal_behind_2004}. We do not address this directly. Instead we assume that an economy with real goods and prices can, for some purposes, be well-represented by a model economy with continuous quantities and prices that can take any real value.}

In the general form of Equation \ref{eqn_pdot_qdot} we can combine prices and quantities into a vector $x$, where
\begin{equation}
x \equiv \begin{pmatrix}p\\\lambda^{-1}q\end{pmatrix}.
\end{equation}
We have pre-multiplied the quantities by the inverse of the coefficient matrix $\lambda$ so that the entire vector $x$ has units of price. With this definition we can write a combined expression for Equations \ref{eqn_pdot_qdot},
\begin{equation}\label{eqn_xdot}
\dot x = \xi(x),
\end{equation}
where
\begin{equation}
\xi(x) \equiv\begin{pmatrix}
\kappa\eta(x)\\
\pi(x)
\end{pmatrix}.
\end{equation}
We shall refer to this as the ``excess demand and price deviation function.'' For an economy with $n$ commodities, Equation \ref{eqn_xdot} provides, for the case of proportional control with cross-effects, an equation for a combined price and quantity vector in a $2n$-dimensional space.

As explained in the Introduction, adjustment mechanisms like Equation \ref{eqn_xdot} are often unstable. However, we also argue that they are unrealistic, and we shall extend Equation \ref{eqn_xdot} in two ways. First, we observe that economic actors will not know the actual excess demand or price deviation reflected in $\xi(x)$. Instead, they will make a guess that will have some error $\epsilon$, so we revise Equation \ref{eqn_xdot} to read
\begin{equation}\label{eqn_xdot_epsilon}
\dot x = \xi(x) + \epsilon.
\end{equation}
Second, we note that Equation \ref{eqn_xdot} expresses the behavior of an active and aware seller who is looking at various market signals---levels of stocks, prices of inputs, sales volumes, etc.---and deciding on that basis to adjust prices, production levels, quantities on shelves or in sales lots, and so on. As we showed in the previous section, and as has been shown in other contexts \cite{simon_behavioral_1984,tversky_judgment_1974,kahneman_maps_2002}, people use heuristics in their economic decision-making. A particular heuristic---doing what others have just done---has been observed in the price-setting behavior of executives in firms \cite{blinder_asking_1998} and in small group settings \cite{ostrom_rules_1994}. We take this heuristic as a motivational example and ask how it might affect Equation \ref{eqn_xdot}.

We propose a simple model in which there are two groups of sellers, group $a$, who follow Equation \ref{eqn_xdot_epsilon}, and group $b$, who set their prices and quantities based on what they see others doing. Because we are motivating the form of the equation rather than deriving it we will work somewhat loosely with a set of difference equations. They are
\begin{subequations}\label{eqn_motivating_2ndOrder}
\begin{align}
\Delta x_{a,t} &= \xi(\bar x_{t-1}) + \bar{\epsilon}_t,\label{eqn_motivating_2ndOrder_a}\\
\Delta x_{b,t} &= \nu \Delta\bar x_{t-1},\label{eqn_motivating_2ndOrder_b}
\end{align}
\end{subequations}
where an overbar indicates an average over all sellers. In Equation \ref{eqn_motivating_2ndOrder_a} we have made the assumption that the function $\xi(x)$ is approximately linear over the range of prices and quantities of different sellers, so that $\overline{\xi(x)}\approx\xi(\bar x)$. The change in the average of $x$, $\Delta \bar x$, is then given by
\begin{equation}
\Delta \bar{x}_t = f_a\xi(\bar x_{t-1}) + f_a\bar{\epsilon}_t + f_b\nu\Delta\bar x_{t-1},
\end{equation}
where $f_a$ is the fraction of all sellers in group $a$ and $f_b$ is the fraction in group $b$, with $f_a + f_b = 1$.
Defining the second-order difference $\Delta^2 \bar x_t \equiv \Delta \bar x_t - \Delta \bar x_{t-1}$, and rearranging, we can rewrite this equation as
\begin{equation}\label{eqn_xddelta}
\Delta^2\bar{x}_t = \frac{f_a}{f_b\nu}\xi(\bar x_{t-1}) + \frac{f_a}{f_b\nu}\bar{\epsilon}_t -\frac{1-f_b\nu}{f_b\nu}\Delta\bar x_t.
\end{equation}
Returning to a continuous-time formulation, and absorbing and redefining factors involving $f_a$, $f_b$, and $\nu$, we propose the following adjustment process,
\begin{equation}\label{eqn_xddot}
\ddot x = \xi(x) - \gamma\cdot\dot x + \epsilon.
\end{equation}
This equation has two features of particular interest. First, like the derivative-control mechanism of \cite{flaschel_dynamic_1997}, it is a second-order process. Second, it has a damping term, $-\gamma\cdot\dot x$. We assume that $\gamma$ is a diagonal matrix with all positive elements.

Comparing Equation \ref{eqn_xddot} to Equation \ref{eqn_xddelta} we see that the damping coefficient $\gamma$ corresponds to $(1 - f_b\nu)/f_b\nu$. Thus in this model damping decreases as either $f_b$ or $\nu$ increases. This shows that the damping is due to group $a$, the active market participants; they move the system toward equilibrium by taking advantage of profitable opportunities. However, they are recurrently given new opportunities for profit by the actions of group $b$.

We note that there are several routes to a second-order t\^atonnement process, and acknowledge that forms other than Equation \ref{eqn_xddot} are possible. For example, the linear damping term could be replaced by a term in which an increase in price leads to an acceleration in the rate of price increases when prices are considered too low (a bull market) and a deceleration in the rate of price increases when prices are considered too high (a bear market). Also, we only included one time lag, while the motivating argument that we proposed is consistent with multiple and overlapping time lags. Furthermore, the model lacks numerous important details, such as overlapping generations with lifetime consumption trajectories, entry into and exit from markets, savings, credit, and an explicit treatment of stocks. Nevertheless, we argue that Equation \ref{eqn_xddot} is more realistic than Equation \ref{eqn_xdot} and, although it is only a modest extension, it has substantively different properties that are worth exploring.

\subsection{Decomposing the excess demand and price deviation function}
Equation \ref{eqn_xddot} can be rewritten in an interesting and compelling way by applying the Helmholtz-Hodge decomposition theorem; a semi-formal proof of the theorem is provided in Appendix \ref{appendix_hh}. This theorem, which is a standard result of differential geometry, states that any continuous vector function defined over a compact space can be written as the sum of the (negative) gradient of a scalar potential $\phi(x)$ and a divergence-free vector field $\alpha(x)$. Therefore,
\begin{equation}\label{eqn_xddot_decomp}
\ddot x = -\nabla\phi(x) + \alpha(x) - \gamma\cdot\dot x + \epsilon,
\end{equation}
where $\nabla\cdot\alpha(x)=0$. Dot-multiplying Equation \ref{eqn_xddot_decomp} on the left with the time rate of change of $x$, $\dot x$, gives
\begin{equation}\label{eqn_energy_dt}
\frac{d}{dt} \left(\frac{1}{2}\dot x^2 + \phi(x)\right) = \dot x\cdot\alpha(x) - \dot x\cdot\gamma\cdot\dot x + \dot x\cdot\epsilon.
\end{equation}
This is an interesting equation. The expression on the left corresponds to the physical energy of a particle moving in a conservative potential. The terms on the right add to or subtract from this ``energy''. However, it is not energy, and to avoid confusion we shall refer to it as ``activity''. The activity has two terms, one measuring how rapidly prices and quantities are changing and the other measuring how far away the system is from an equilibrium point.

First, suppose that both $\alpha(x)$ and $\epsilon$ are zero. In this case Equation \ref{eqn_energy_dt} becomes
\begin{equation}
\frac{d}{dt} \left(\frac{1}{2}\dot x^2 + \phi(x)\right) = - \dot x\cdot\gamma\cdot\dot x.
\end{equation}
The right-hand side of the equation is negative definite, so it continuously reduces the left-hand side until $\dot x=0$ at a stable point where $\phi(x)$ is at a minimum. This shows that the term $\dot x\cdot\gamma\cdot\dot x$ is a damping term that serves to reduce the activity in the system and bring it to a stable equilibrium. However, while the rational expectations hypothesis implies that $\epsilon$ should be zero, it is unrealistic to expect that $\alpha(x)$ is zero. If it were, then the excess demand and price deviation function $\xi(x)$ would be given solely by the gradient of the potential, so that $\xi_j(x) = -\partial_j\phi(x)$, where $\partial_j$ is shorthand for the partial derivative with respect to $x_j$. In this case, the Jacobian matrix $\partial_i\xi_j(x)$ is equal to $-\partial_i\partial_j\phi(x)$ and is therefore symmetric. This would mean that for every pair of commodities $(i,j)$, commodity $i$ responds to changes in commodity $j$ in exactly the same way that commodity $j$ responds to changes in commodity $i$. Except in the case of exact substitutes this is unlikely to occur. We therefore assume that $\alpha(x)$ is generically nonzero. We also assume that $\epsilon$ might be of either sign, or zero.

\subsection{Limit cycles}
Suppose that prices and quantities enter a limit cycle, a closed path $z(t)$ that repeats endlessly. In this case we can integrate over the time it takes to execute the loop. The net change of activity is zero, because it is an exact integral over time and the beginning and ending points have the same value. Therefore, for a repeating loop,
\begin{equation}\label{eqn_limit_cycle_orig}
\oint_z dx\cdot\alpha(x) - \oint_z dx\cdot\gamma\cdot\dot x + \oint_z dx \cdot\epsilon = 0.
\end{equation}
The second of these terms is negative definite, and represents a path-dependent dissipative force, $-D[z]$. The third term might be positive, negative, or zero; it represents the average bias $B[z]$ in the assessments of sellers of the state of the market.

The first term can be rewritten using Stokes' theorem, another standard result from differential geometry. In arbitrary dimensions we represent this as an integral over a two-form $d\omega$, where
\begin{equation}
d\omega = \partial_j\alpha_i(x) dx^j \wedge dx^i.
\end{equation}
From Stokes' theorem we have
\begin{equation}
\int_{S(z)} d\omega = \oint_z \omega = \oint_z \alpha_i(x)dx^i = \oint dx\cdot\alpha(x).
\end{equation}
In the left-most integral, $S(z)$ is a surface bounded by the closed path $z$. Because the wedge product $dx^j\wedge dx^i$ is anti-symmetric, the two-form $d\omega$ consists of the anti-symmetric parts of the Jacobian matrix for $\alpha(x)$,
\begin{equation}
d\omega = \frac{1}{2} \left( \partial_i\alpha_j(x) - \partial_j\alpha_i(x) \right) dx^i\wedge dx^j.
\end{equation}
However, this is also equal to the anti-symmetric part of the Jacobian matrix of the full excess demand and price deviation function $\xi(x)$, because the rest of the Jacobian is the second partial derivative matrix of the scalar potential, which (as already noted above), is symmetric. Therefore,
\begin{equation}\label{eqn_domega}
d\omega = \frac{1}{2} \left( \partial_i\xi_j(x) - \partial_j\xi_i(x) \right) dx^i\wedge dx^j.
\end{equation}

Combining Equation \ref{eqn_domega} with Equation \ref{eqn_limit_cycle_orig}, and using the notation for the dissipative and expectation bias terms, we can write
\begin{equation}
\frac{1}{2} \oint_{S(z)} \left( \partial_i\xi_j(x) - \partial_j\xi_i(x) \right) dx^i\wedge dx^j - D[z] + B[z] = 0.
\end{equation}
What this equation says is that prices and quantities can endlessly cycle if the dissipative forces arising from the profit-maximizing behavior of a segment of sellers is balanced by either a positive bias in their assessment of the state of the market or an asymmetry in the Jacobian of the excess demand and price deviation function. We interpret the first term in this equation to mean that those sellers who are actively watching for profit opportunities are repeatedly offered ``favorable surprises'' \cite{fisher_disequilibrium_1983} due to asymmetries in the way that prices and quantities of one commodity responds to changes in another commodity. The asymmetry is manifest as a recurring favorable surprise because of the second-order t\^atonnement process. It is a non-conservative term that can either add to or subtract from activity.

\subsection{Asymptotic behavior of trajectories: equilibrium or circulation}
While limit cycles are possible, they arise in models with specific forms \cite{li_bendixsons_1993}, so they are unlikely to occur for an arbitrary excess demand and price deviation function. Nevertheless, any trajectory must either end at rest at a minimum of the scalar potential or circulate around a region where the excess demand and price deviation function has a nonzero curl. To see this, we integrate Equation \ref{eqn_energy_dt} over a trajectory $z(t)$ that does not close. In this case,
\begin{equation}\label{eqn_deltaA_general}
\Delta A(t) = \int_z dx\cdot\alpha(x) - \int_z dx\cdot\gamma\cdot\dot{x} + \int_z dx\cdot\epsilon,
\end{equation}
where $\Delta A(t)$ is the change in activity between the endpoints of the path. The second term is always negative and always subtracts from activity as long as prices or quantities are changing. It will therefore bring activity to a local minimum of the scalar potential with $\dot x = 0$ unless offset by the first or third terms. But the first or third terms must perpetually add to the activity. For the third term this can only happen if sellers maintain positively biased expectations indefinitely into the future. We argue that this extreme departure from the rational expectations model is unlikely, and set the long-run value of $\epsilon$ to zero. The burden then falls on the first term to perpetually add to activity.

The indefinitely extended path $z(t)$ has endpoints that differ from one another. There is another path $z'(t)$ that is the shortest path between the same endpoints. The difference in the integral over $\alpha(x)$ between the two paths is
\begin{equation}
\int_z dx\cdot\alpha(x) - \int_{z'} dx\cdot\alpha(x) = \oint_{z-z'} dx\cdot\alpha(x),
\end{equation}
where $z-z'$ is the closed path that first traverses $z(t)$ in a positive direction and then $z'(t)$ in a negative direction. If the path does not enclose a region in which the curl of $\alpha(x)$, and therefore the curl of $\xi(x)$, is non-zero, then the integral vanishes. Accordingly, in this case the first term in Equation \ref{eqn_deltaA_general} can be replaced by $\int_{z'} dx\cdot\alpha(x)$, and
\begin{equation}\label{eqn_deltaA_nocurl}
\Delta A(t) = \int_{z'} dx\cdot\alpha(x) - \int_z dx\cdot\gamma\cdot\dot{x}.
\end{equation}
But in a compact space the path $z'(t)$, which is the shortest path between the endpoints of $z(t)$, cannot extend indefinitely, and therefore the first term is bounded. Because the second term is not bounded it must eventually bring the system to a halt. We therefore conclude that price and quantity trajectories either terminate at a local minimum of the scalar potential or circulate endlessly around an area where the curl of the excess demand and price deviation function is non-zero.\footnote{We note that the system behaves as though there were a topological defect associated with the breaking of an exchange symmetry between goods. However, a true topological defect would require a theory of the excess demand and price deviation function that has a global exchange symmetry. That does not seem reasonable---some types of goods are fundamentally different from others. However, it might accurately describe large segments of the economy. For example, advertising, market share, or an appeal to store loyalty can make one otherwise indistinguishable product stand out from the others.}

\subsection{Stability against perturbations in a subspace of substitutable commodities}
We now show that the system is stable against a perturbation in any subspace where the dynamics are described entirely by a scalar potential. An important example is a subspace of goods that are exact (or almost exact) substitutes. As the review of price-setting behavior showed, even homogeneous commodities can have different prices because they appeal to consumers for some reason other than the quality and features of the physical commodity itself. We return to this issue in the next section, but for the moment consider the case of a truly competitive market for a homogeneous commodity.

In general, for a small deviation $\delta x$ near an equilibrium point $x^*$, the excess demand and price deviation function is approximately
\begin{equation}
\xi(x^*+\delta x)\approx \delta x \cdot \nabla\xi(x^*).
\end{equation}
The gradient of the excess demand and price deviation function is the Jacobian matrix with elements $(\nabla\xi(x^*))_{i,j} = \partial_i\xi_j(x^*)$. Within a subset of the full commodity space $\Sigma$ where all commodities are exact substitutes, the Jacobian matrix is highly regular, and the values $\gamma_i$ for the (diagonal) damping coefficient matrix will all be the same, $\gamma_i = \gamma$. Supposing that only the excess demand portion of the excess demand and price deviation function is changing, and solely in response to a change in price, the Jacobian matrix has elements
\begin{equation}
\partial_i\xi_j(x^*)=\left\lbrace \begin{matrix}
-a(x^*), & i=j \\ b(x^*), & i\ne j
\end{matrix} \right., \mbox{ where } i,j\in\Sigma.
\end{equation}
That is, we suppose that the own-price response for each commodity (or brand) is negative---so that a rise in price depresses demand---while the cross-price response is positive---so that a rise in a different, but substitutable commodity or brand increases demand. Within this subspace, and to second order near the equilibrium, the excess demand and price deviation function can be written in terms of a scalar potential
\begin{equation}
\phi(x^*+\delta x)\approx \phi(x^*) + \frac{1}{2}a\sum_{i\in\Sigma} \delta x_i^2 - b\sum_{\stackrel{j>i}{i,j\in\Sigma}} \delta x_i\delta x_j.
\end{equation}
In this case the dynamic equation for $\delta x$ is
\begin{equation}
\frac{d}{dt} \left(\frac{1}{2}\dot {\delta x}^2 + \phi(x^*+\delta x)\right) = - \gamma\dot {\delta x}^2 + \int_{\delta x}dx\cdot\epsilon.
\end{equation}
This equation shows that continued optimism, in the form of a positively-biased expectation $\epsilon$, can keep the system away from equilibrium. However, it seems reasonable in a highly competitive market with many near substitutes that the rational expectations value $\epsilon = 0$ would hold. In that case the disturbance would decay to zero at a rate determined by $\gamma$. This behavior is consistent with the decay portion of the trajectory of gasoline prices in Vancouver observed by Eckert and West \cite{eckert_retail_2004}: more or less weekly, some retailers start a rapid but small surge in prices followed by an uneven and more gradual decay.\footnote{The initial surge is not explained by the theory in this paper; it is a price ``shock'' that is internal to the system and possibly peculiar to the retail gasoline market, which can only be explained by a behavioral model.}

This result---that disturbances into a subspace of substitutes decay---raises the possibility that a trajectory through the commodity space could, for the most part, wander over a sub-manifold delimited by clusters of substitutable commodities. Any deviation in the direction of a subspace of substitutable commodities would decay, bringing the system back to the sub-manifold. We note incidentally that if this describes real systems to a sufficient degree of accuracy then it provides a justification for the common practice of aggregating over similar commodities.

\vspace{6pt}

The arguments in this section suggest, but do not prove, that the markets described by the model are globally stable but locally unstable. Trajectories approach local minima of the scalar potential and then circulate around one or more minima, driven by asymmetries in the way sellers and buyers respond to changes in quantities and prices of different commodities.

\section{Conjecture: Stability through Degeneracy}
In the previous section we showed that deviations into a subspace of perfectly substitutable goods decay. More generally, but also much more difficult to demonstrate, the system could be stable against deviations into sets of \emph{degenerate} commodities, in the sense that the term is used in complex systems theory. Originating in neurobiology \cite{tononi_measures_1999}, degeneracy describes systems in which subsystems can take over each others' functions, even though they are not completely identical; it has been proposed as a design principle leading to robustness and steady evolution in complex systems \cite{whitacre_degeneracy:_2010}. We suggest that degeneracy is likely in economic systems. In a well-known paper, Stigler and Becker \cite{stigler_gustibus_1977} demonstrate that apparently unstable and changing consumption choices can be reconciled with a theory of stable preferences. Their demonstration relies on two propositions: that people combine market goods with their own labor and time to produce internal ``commodities'', and that people build up various ``capitals'' within themselves through their actions and experiences. The list of internal commodities--self-esteem, physical fitness, intellectual stimulation, etc.---is relatively stable and unchanging. The way in which people supply those commodities depends on the capitals that they have built up, and hence on their life-history. We consider only the first proposition, that people seek to satisfy a relatively stable set of internal commodities. This conjures an image of observed demands for physical goods as a fog through which demands for the commodities we truly desire dimly shine, an image that seems more consistent with observed consumer purchasing behavior than with the idealized model of a consumer continually optimizing in the world of physical goods. Mathematically, there is a space of (shadow) prices and quantities of unseen but fundamental commodities $y$ that are met by combining personal efforts with physical goods $x$. Denoting the excess demand and price deviation function within this fundamental commodity space by $\chi(y)$, the excess demand and price deviation function for the space of physical commodities is $\xi_i(x) = \sum_{j=1}^{2m} \langle({\partial y_j}/{\partial x_i})\chi_j(y)\rangle$, where $m$ is the number of internal commodities, and the average $\langle\cdot\rangle$ is taken over all consumers.

If there were only one type of consumer---that is, if all consumers agreed on how to meet their demand for internal commodities---then the matrix ${\partial y_j}/{\partial x_i}$ would be the same for everyone. Because we presume the dimensionality of the internal commodity space $2m$ to be much less than the dimensionality of the space of physical goods $2n$, this is a highly degenerate matrix in the mathematical sense. There will be a $2(n-m)$-dimensional subspace over which the physical excess demand and price deviation function is (locally) invariant. Consumers are not homogeneous and so they will not have the same matrix ${\partial y_j}/{\partial x_i}$. Nevertheless, to the extent that there is some convergence in consumer behavior we expect that degeneracy in the systems-theoretical sense---diverse physical goods meeting the same internal needs---will give rise to locally invariant subspaces of the excess demand and price deviation function for physical goods. As shown earlier, deviations from equilibrium within a subspace consisting of substitutable goods return to equilibrium.

This leads to the following conjecture: Market goods are highly degenerate, in the systems-theoretical sense, in that different combinations of goods fulfill the same function of satisfying demands for internal ``commodities''. As a result, at the level of the economy as a whole there is on average a local and approximate invariance between exchanges of groups of goods, and the invariant subspace is of high dimension. Further, deviations within the locally invariant subspace decay more or less quickly, and do not diverge or enter into large-amplitude cycles. As a consequence, prices and quantities of goods remain bound to a relatively low-dimensional manifold within the full space of possible prices and quantities.

\section{Conclusion}
If we convince the reader on no other point, we hope at least to persuade him or her that there is value in looking again, and more closely, at dynamic economic theories. As deep problems with dynamic theories emerged, General Equilibrium theory, and the efficient markets hypothesis on which it rests, gained popularity. But the efficient market hypothesis is known to fail empirical tests for financial markets \cite{kristoufek_fractal_2012}, and is inconsistent with observed behavior of firms, retailers, and consumers.

The approach we took in this paper is to build a highly aggregate and idealized model of the economy, a fairly modest---yet substantively different---variation on existing models of price and quantity-adjustment. While lacking some important features of real economies, the model is very general, permitting almost any functional form for the excess demand and price deviation function. We showed that the model exhibits convergence toward local minima of a scalar potential, with possibly endless cycling around one or more minimum. The cycles are generated by asymmetries in the response of sellers and buyers to changes in price or quantity of different commodities; in the model the cycles surround regions of commodity space where the curl of the excess demand and price deviation function does not vanish. These interesting features of our somewhat naive aggregate model suggests that research into more explicit, and more empirically-based, aggregation procedures could be fruitful. For example, network models \cite{barabasi_scale-free_2009} or fractal market models \cite{kristoufek_fractal_2012} could generate interesting functional forms and provide insights into the operation of real markets.

The model in this paper features convergence in subspaces of identical commodities. Thus, if superficially similar commodities---such as sticks of butter---really were identical in fact, then price or quantity deviations into the space of such commodities will decay. If true, this provides a justification for the common practice of aggregating over similar commodities. However, as discussed in the section on empirical observations of seller and consumer behavior, prices of functionally identical commodities can be different from each other as a result of advertising, while consumers are strongly influenced by brands. A perhaps more promising avenue to incorporating product substitutability into a market model is the theory of Stigler and Becker \cite{stigler_gustibus_1977} that people purchase goods in diverse combinations in order to satisfy their desire for a stable set of internal commodities. We conjectured that if there is some convergence across consumers on how goods can be combined to satisfy internal needs and wants, then price and quantity trajectories should mainly lie on a low-dimensional manifold within the full commodity space. This would include out-of-equilibrium paths, and so it would be larger than the equilibrium manifold \cite{balasko_equilibrium_2009}, but would be quite a bit smaller than the full space.

\newpage\bibliography{price_quantity_diffgeom}

\newpage\appendix
\section{The Helmholtz-Hodge Decomposition Theorem}\label{appendix_hh}
In this appendix we provide a semi-formal proof of the Helmholtz-Hodge decomposition theorem. That is, we motivate all of the steps of the proof, but do not reach the level of rigor required by a mathematician.\footnote{The proof is a straightforward generalization to arbitrary dimensions of a classical proof for three dimensions.}

We shall show that a continuous vector field $v(x)$ in a $D$-dimensional space can be written
\begin{equation}
v(x) = -\nabla \phi(x) + \alpha(x),
\end{equation}
where $\phi(x)$ is a scalar potential and $\alpha(x)$ is a divergence-free vector field, $\nabla\cdot\alpha(x) = 0$. We argue by construction, first defining a vector field
\begin{equation}\label{eqn_appx_psidef}
\psi(x) \equiv -\frac{\Gamma(D/2)}{2\pi^{D/2}(D-2)} \int_V d^Dx'\thinspace \frac{v(x')}{R^{D-2}},
\end{equation}
where $R\equiv |x-x'|$. We also define the scalar and vector fields
\begin{subequations}\label{eqn_appx_phiadefs}
\begin{align}
\phi(x) &\equiv -\nabla\cdot\psi(x),\\
\alpha(x) &\equiv (\nabla^2 - \nabla\nabla\cdot)\psi(x).\label{eqn_appx_phiadefs_b}
\end{align}
\end{subequations}
Note that, because $\nabla^2$ and $\nabla\cdot$ commute, Equation \ref{eqn_appx_phiadefs_b} implies that $\nabla\cdot\alpha(x)=0$, as required.

Below, we show that
\begin{equation}\label{eqn_appx_greenfunction}
\nabla^2 \psi(x) = v(x).
\end{equation}
Assuming that this equation is true, we find
\begin{equation}
(\nabla^2 - \nabla\nabla\cdot)\psi(x) = v(x) - \nabla\nabla\cdot\psi(x).
\end{equation}
Combined with Equations \ref{eqn_appx_phiadefs}, this equation can be rearranged to show that
\begin{equation}
v(x) = -\nabla\phi(x) + \alpha(x),
\end{equation}
which is the result we seek. The proof therefore hinges on Equation \ref{eqn_appx_greenfunction}, which we prove in the remainder of this appendix.

Equation \ref{eqn_appx_greenfunction} would have us compute the Laplacian of $\psi(x)$. Because the Laplacian acts on the unprimed variables it can be moved under the integral in Equation \ref{eqn_appx_psidef} and applied to $R$,
\begin{equation}\label{eqn_appx_greenfunction_explicit}
\nabla^2\psi(x) \equiv -\frac{\Gamma(D/2)}{2\pi^{D/2}(D-2)} \int_V d^Dx'\thinspace v(x')\nabla^2\frac{1}{R^{D-2}}.
\end{equation}
The core of the calculation is the Laplacian of $1/R^{D-2}$. Expressed explicitly in terms of the variables $\{x_i\}$, $1/R^{D-2}$ is
\begin{equation}
\frac{1}{R^{D-2}} = \left[\sum_{i=1}^D (x_i - x'_i)^2\right]^{-\frac{D-2}2}.
\end{equation}
The first partial derivative of this expression with respect to variable $x_j$ is
\begin{equation}\label{eqn_appx_gradinvR}
\frac{\partial}{\partial x_j}\left[\sum_{i=1}^D (x_i - x'_i)^2\right]^{-\frac{D-2}{2}} = -(D-2)\left[\sum_{i=1}^D (x_i - x'_i)^2\right]^{-\frac{D}{2}} (x_j - x'_j).
\end{equation}
We will need this expression later. Noting that the unit vector $\hat{r}$ in the direction of $x-x'$ has components
\begin{equation}
\hat{r}_i = \frac{x_i - x'_i}{\sqrt{\sum_{i=1}^D (x_i - x'_i)^2}}=\frac{x_i-x'_i}{R},
\end{equation}
Equation \ref{eqn_appx_gradinvR} shows that the gradient of $1/R^{D-2}$ is
\begin{equation}\label{eqn_appx_gradinvR_pretty}
\nabla \frac{1}{R^{D-2}} = -\frac{D-2}{R^{D-1}}\hat{r}.
\end{equation}

The Laplacian of $1/R^{D-2}$ is found by taking a further partial derivative of Equation \ref{eqn_appx_gradinvR} with respect to $x_j$ and summing over all values of $j$, from 1 to $D$. Taking the partial derivative results in
\begin{equation}
-(D-2)\frac{\partial}{\partial x_j}\left[\sum_{i=1}^D (x_i - x'_i)^2\right]^{-\frac{D}{2}} (x_j - x'_j) = -(D-2)\left[\sum_{i=1}^D (x_i - x'_i)^2\right]^{-\frac{D}{2}}+D(D-2)\left[\sum_{i=1}^D (x_i - x'_i)^2\right]^{-\frac{D}{2}-1}(x_j - x'_j)^2.
\end{equation}
Thus the Laplacian is given by
\begin{equation}
\nabla^2\frac{1}{R^{D-2}} = -(D-2)\left[\sum_{i=1}^D (x_i - x'_i)^2\right]^{-\frac{D}{2}} \left[\sum_{j=1}^D 1-D\frac{\sum_{j=1}^D(x_j - x'_j)^2}{\sum_{i=1}^D (x_i - x'_i)^2}\right].
\end{equation}
Because $\sum_{j=1}^D 1 = D$ and the ratio in the second set of brackets is equal to one (except perhaps when $x=x'$, where it is undefined), this becomes
\begin{equation}
\nabla^2\frac{1}{R^{D-2}} = -(D-2)\left[\sum_{i=1}^D (x_i - x'_i)^2\right]^{-\frac{D}{2}} (D-D) = 0 \mbox{  when } x\neq x'.
\end{equation}
That is, the Laplacian of $1/R^{D-2}$ is equal to zero except perhaps when $x=x'$. This explains the result we seek to prove, Equation \ref{eqn_appx_greenfunction}: the Laplacian of $1/R^{D-2}$ is a representation of the Dirac delta function in $D$ dimensions.

To show that this is, indeed, the case, we replace the integral in Equation \ref{eqn_appx_greenfunction_explicit} with one that only extends to a very small hyper-ball of radius $\epsilon$, where $\epsilon$ is small enough that we can consider $v(x)$ to be approximately constant. We can do this---in fact, we can do it for a ball as small as we choose---because we know that the integrand is zero everywhere except perhaps when $x=x'$. Over this small volume we can write,
\begin{equation}
\nabla^2\psi(x) \approx -\frac{\Gamma(D/2)}{2\pi^{D/2}(D-2)} v(x) \int_{B_\epsilon} d^Dx'\thinspace \nabla^2\frac{1}{R^{D-2}},
\end{equation}
where $B_\epsilon$ is the hyper-ball of radius $\epsilon$ in $D$ dimensions. The integral of the Laplacian of $1/R^{D-2}$ over the volume of the hyper-ball gives an integral of normal component of the gradient, $\hat{n}\cdot\nabla(1/R^{D-2})$ over the $D-1$ dimensional hypersphere $\partial B_\epsilon$ that is the boundary of $B_\epsilon$. That is,
\begin{equation}
\nabla^2\psi(x) \approx -\frac{\Gamma(D/2)}{2\pi^{D/2}(D-2)} v(x) \int_{\partial B_\epsilon} dS \thinspace \hat{n}\cdot\nabla\frac{1}{R^{D-2}},
\end{equation}
But we have an expression for the gradient of $1/R^{D-2}$ in Equation \ref{eqn_appx_gradinvR_pretty}. Substituting this expression, and using the fact that $R=\epsilon$ and $\hat{n} = \hat{r}$ on the surface of the hyper-ball, we find that
\begin{equation}
\nabla^2\psi(x) \approx \frac{\Gamma(D/2)}{2\pi^{D/2}} v(x) \frac{S(\partial B_\epsilon)}{\epsilon^{D-1}},
\end{equation}
where $S(\partial B_\epsilon)$ is the surface area of the $D-1$ hypersphere of radius $\epsilon$.
It is a standard result that $S(\partial B_\epsilon) = \left[2\pi^{D/2}/\Gamma(D/2)\right]\epsilon^{D-1}$ (e.g., \cite{huber_gamma_1982}), so this approximate equality reduces to
\begin{equation}
\nabla^2\psi(x) \approx v(x).
\end{equation}
Because this is independent of $\epsilon$, in the limit that $\epsilon$ goes to zero we can replace the approximation sign with an equality,
\begin{equation}
\nabla^2\psi(x) = v(x),
\end{equation}
which is the result we sought to prove.

\end{document}